\documentclass[lettersize,journal]{IEEEtran}
\usepackage{amsmath,amsfonts}
\usepackage{algorithmic}
\usepackage{array}
\usepackage[caption=false,font=normalsize,labelfont=sf,textfont=sf]{subfig}
\usepackage{textcomp}
\usepackage{stfloats}
\usepackage{multirow}
\usepackage{tabularx}
\usepackage{float}
\usepackage{multirow}
\usepackage{makecell}
\usepackage{url}
\usepackage{booktabs}
\usepackage{verbatim}
\usepackage{graphicx}
\usepackage{xcolor}
\def\BibTeX{{\rm B\kern-.05em{\sc i\kern-.025em b}\kern-.08em
		T\kern-.1667em\lower.7ex\hbox{E}\kern-.125emX}}
\usepackage{balance}
\begin{document}

	\title{A Hybrid Model-Assisted Approach for Path Loss Prediction in Suburban Scenarios}
	
	\author{Chenlong Wang,
    ~\IEEEmembership{Student Member,~IEEE}, \and
		Bo Ai,~\IEEEmembership{Fellow,~IEEE,} \and
        Ruiming Chen, \and
		Ruisi He,~\IEEEmembership{Senior Member,~IEEE,} \and
        Mi Yang,~\IEEEmembership{Member,~IEEE}, \and
		Yuxin Zhang,~\IEEEmembership{Student Member,~IEEE,} \and
        Weirong Liu, \and
        Liu Liu,~\IEEEmembership{Member,~IEEE}

		\thanks{
			
			Chenlong Wang, Bo Ai, Ruiming Chen, Ruisi He, Mi Yang, Yuxin Zhang, Weirong Liu and Liu Liu are with the State Key Laboratory of Advanced Rail Autonomous Operation, the School of Electronics and Information Engineering, and the Frontiers Science Center for Smart High-speed Railway System, Beijing Jiaotong University, Beijing 100044, China (email: wang-chenlong@bjtu.edu.cn; aibo@ieee.org; 21211059@bjtu.edu.cn; ruisi.he@bjtu.edu.cn; myang@bjtu.edu.cn; yx-zhang@bjtu.edu.cn; bjtulwr@bjtu.edu.cn; liuliu@bjtu.edu.cn).
		}
		
	}
	
	\markboth{Journal of \LaTeX\ Class Files,~Vol.~14, No.~8, August~2021}
	{Shell \MakeLowercase{\textit{et al.}}: A Sample Article Using IEEEtran.cls for IEEE Journals}

	\maketitle
	
	\begin{abstract}
        Accurate path loss prediction is crucial for wireless network planning and optimization in suburban environments with complex terrain variation and diverse land cover. This paper proposes a model-assisted hybrid path loss prediction method that introduces an environment-adaptive compensation on top of the classic close-in free-space reference distance (CI) path loss model. By jointly predicting a link-specific effective path loss exponent and a compensation term, the proposed approach dynamically adjusts the empirical trend. To improve the effectiveness of environmental representation, three environmental image organization schemes are constructed and evaluated. Experiments on measurement data collected in Pingtan Island show that the proposed method outperforms the CI model and a conventional model-assisted baseline, achieving a root mean square error of 4.04 dB on the test set.
	\end{abstract}
	
	\begin{IEEEkeywords}
		Path loss prediction, model-assisted learning, channel prediction, deep learning.
	\end{IEEEkeywords}

	\section{Introduction}
	
	Accurate path loss prediction is essential for wireless network planning and optimization, particularly in suburban environments where terrain undulation, building distribution, and heterogeneous land cover introduce strong environment-dependent propagation characteristics. As wireless communications continue to evolve from fifth-generation (5G) to sixth-generation (6G) systems, propagation scenarios are becoming increasingly diverse, and conventional models are no longer sufficient for site-specific deployment evaluation, coverage prediction, and link budget analysis \cite{9103348}. Therefore, developing path loss prediction methods that achieve both high accuracy and practical efficiency has become increasingly important.
    
    Existing path loss prediction approaches still involve a significant tradeoff among prediction accuracy, computational complexity, and generalization capability \cite{10839242}. Traditional empirical models are easy to deploy and physically intuitive, but their fixed analytical forms make it difficult to characterize nonlinear attenuation caused by local terrain variations and environmental heterogeneity. Deterministic methods can provide higher-fidelity predictions by explicitly modeling propagation mechanisms, yet they usually require detailed geographic reconstruction and considerable computational cost, which limits their scalability in practical applications \cite{11363248}. In recent years, data-driven methods have become an important direction for path loss modeling due to their strong nonlinear fitting ability \cite{10949588, 11141746}. \cite{kwon2025machine} incorporated diffraction parameter and morphology ratio features into machine learning models to enhance the representation of propagation characteristics. However, purely learning-based approaches often suffer from limited physical interpretability, strong dependence on labeled measurement data, and poor robustness when training data are insufficient or environmental conditions vary significantly \cite{11389917, 11373296}.
    
    To balance physical consistency and predictive flexibility, model-assisted approaches, which combine conventional empirical propagation models with neural networks, have gradually emerged as an effective hybrid modeling paradigm \cite{8950164, 9771737, 11310437}. In this framework, the empirical model provides a physically meaningful baseline prediction, while the neural network learns environment-related compensation from data. Although this strategy is more interpretable than purely data-driven methods, most existing model-assisted approaches still rely on relatively shallow environmental representations and simple additive correction structures \cite{11202166,10745208}. As a result, the interaction between physical priors and learned features remains insufficient, which limits their prediction performance in complex suburban scenarios.

    The main contributions of this work are threefold. First, three different environmental image construction schemes are designed and compared for suburban path loss prediction, and the influence of environmental representation scale on model performance is systematically analyzed. Second, a model-assisted path loss prediction method is developed by integrating empirical modeling with multimodal deep feature extraction, in which both feature fusion and environmental compensation are unified within the same framework to improve prediction accuracy and physical rationality. Third, extensive validation using measured data from Pingtan demonstrates that the proposed method outperforms the considered baseline methods.

    \section{Dataset Construction}

\begin{figure}[!t]
    	\centering
    	\includegraphics[width=3.5in]{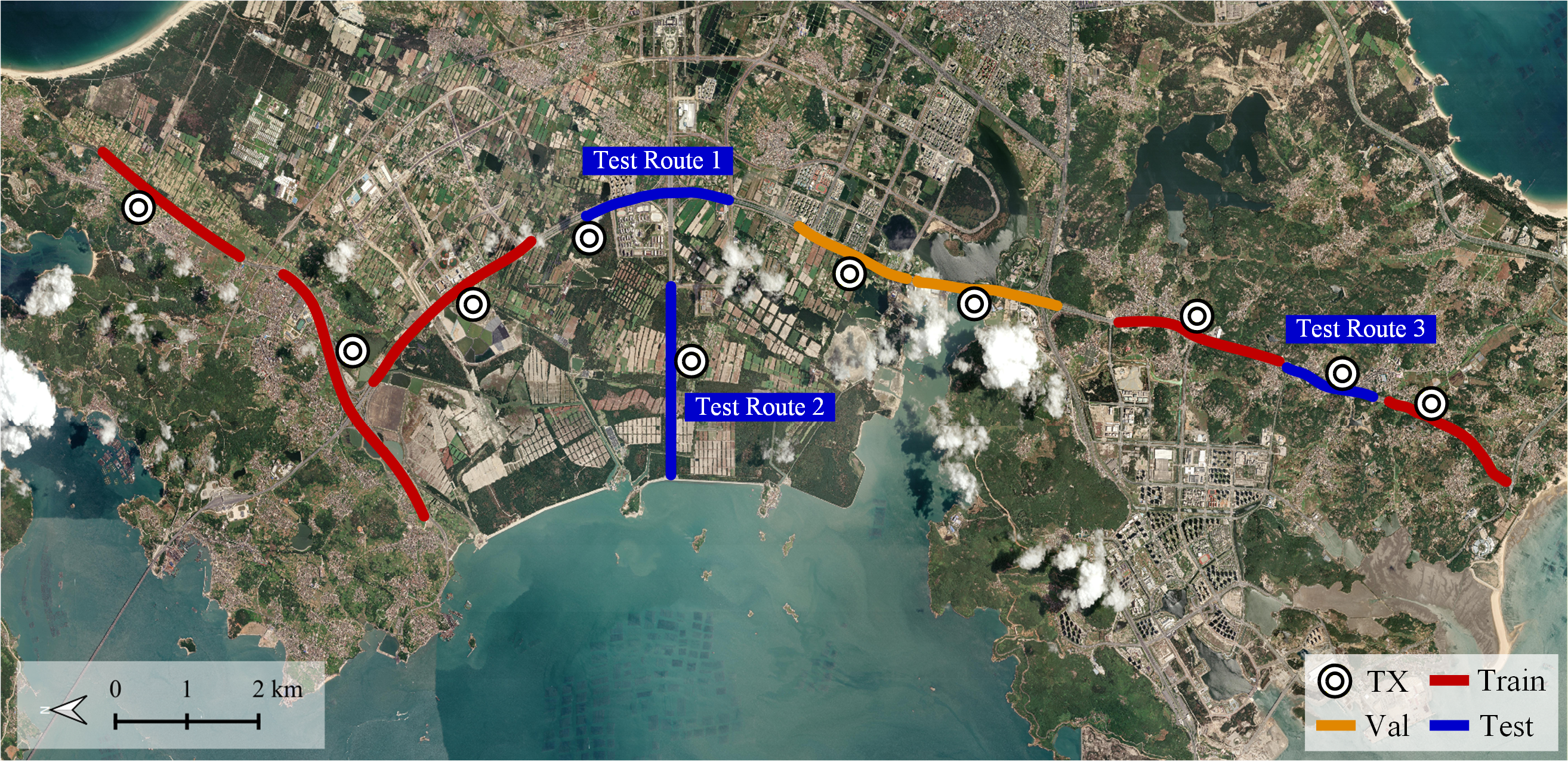}
    	\caption{Overview of the measurement data distribution in Pingtan, Fujian Province, China. The red routes indicate the training set, the yellow routes indicate the validation set, and the blue routes indicate the test set. }
    	\label{overview}
    \end{figure}

	The dataset used in this work is constructed from channel measurement data collected in Pingtan, Fujian Province, China. The measurement system mainly consists of a vector signal transmitter (Tx), a vector signal receiver (Rx), a power amplifier, antennas, and a reference clock. The campaign was conducted at a carrier frequency of 1210 MHz with a 20 MHz bandwidth and a sampling rate of 61.44 MHz, using an omnidirectional antenna. Along each measurement route, the transmitter was in most cases fixed at the top of a hill, while the receiver was integrated into a mobile measurement vehicle traveling at approximately 30 km/h. As shown in Fig. \ref{overview}, the dataset is split by measurement route to better assess the generalization capability of the model. The numbers of samples in the training, validation, and test sets are 9285, 2207, and 4525, respectively. 
    
    Considering that path loss in suburban scenarios is affected by both land-cover characteristics and terrain variations, satellite images and elevation maps are used as multimodal environmental inputs. Both types of data are provided by Geovis Technology Co., Ltd. \cite{geovis}. The satellite images and elevation maps have spatial resolutions of approximately 1 m and 35 m, respectively. During preprocessing, the coordinate reference systems are unified using the open-source QGIS Geographic Information System \cite{QGIS_software}, and both layers are resampled and rendered using the same geospatial crop extent, rotation angle, and target output size, thereby generating spatially aligned multi-channel input matrices.
    
    \begin{figure}[!t]
    	\centering
    	\includegraphics[width=3in]{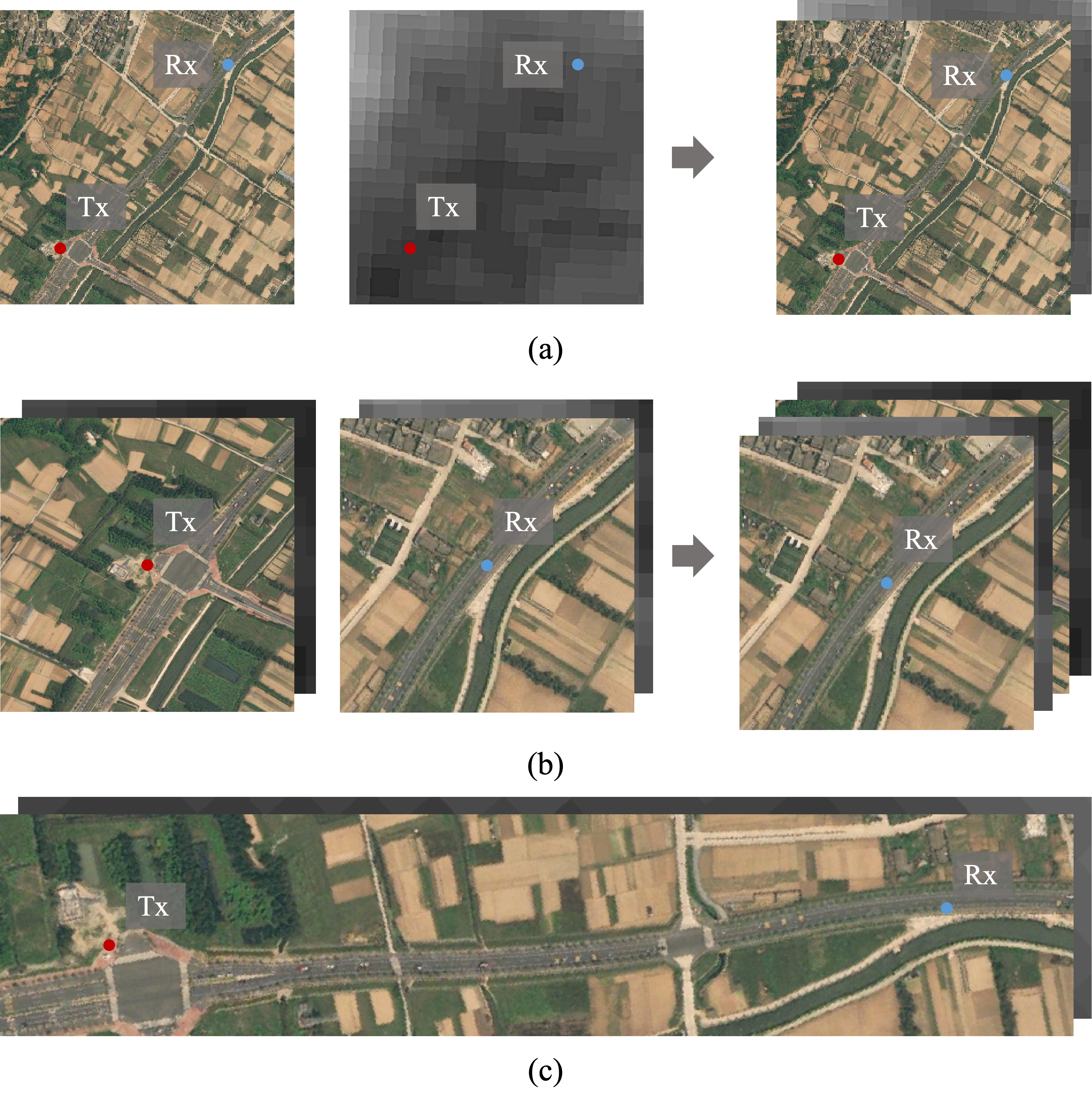}
    	\caption{Illustration of the three environmental image input formats. (a) Resize format. (b) Stacksize format. (c) Fullsize format.}
    	\label{input}
    \end{figure}
    
    To investigate the impact of environmental representation on model performance, three input formats are constructed, namely Resize, Stacksize, and Fullsize, as illustrated in Fig. \ref{input}. Resize is shown in Fig. \ref{input}(a), where the image size is fixed to $256\times256$ pixels. The transmitter is placed at the lower-left corner and the receiver at the upper-right corner. The stretching ratio is adjusted according to the Tx--Rx distance such that the separation always occupies 3/5 of the image diagonal, thereby keeping a consistent scale for different Tx--Rx links. Stacksize is shown in Fig. \ref{input}(b), where two $256\times256$ local patches centered at the transmitter and receiver are extracted and concatenated along the channel dimension to emphasize local details around the transmitter and receiver. Fullsize is shown in Fig. \ref{input}(c). The midpoint of the Tx--Rx segment is used as the image center, and the image height is fixed to 160 pixels. The image width is set to the Tx--Rx distance converted to pixels plus 160 pixels to preserve the link-centered environment.
	
	\section{Model Design and Implementation}
	
\begin{figure}[!t]
    	\centering
    	\includegraphics[width=\columnwidth]{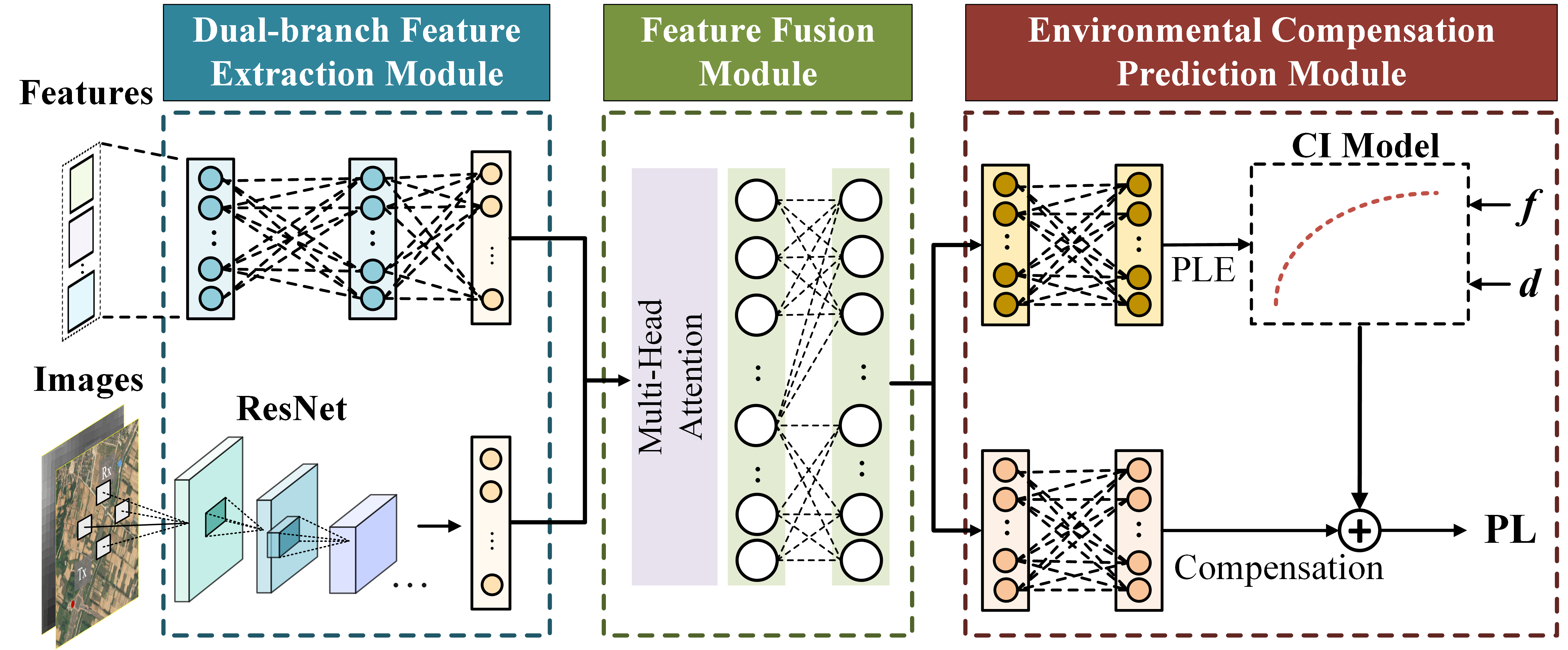}
    	\caption{Overview of the proposed model-assisted hybrid architecture, including the dual-branch feature extraction module, the MHSA-based feature fusion module, and the environmental compensation prediction module.}
    	\label{model}
    \end{figure}

	We propose a model-assisted hybrid path loss prediction approach that combines the physical prior of the CI model with the nonlinear representation capability of deep neural networks. As shown in Fig. \ref{model}, the architecture consists of a dual-branch feature extraction module, a feature fusion module, and an environmental compensation prediction module.

	In the dual-branch feature extraction module, the image branch takes the multichannel environmental image as input and uses a ResNet 50 backbone to learn propagation information from the satellite image and elevation map, such as land-cover distribution and terrain variation. A linear projection layer is appended to obtain a 256-dimensional image feature vector. In parallel, the system branch uses six scalar inputs that are directly related to path loss, including the three-dimensional Tx--Rx distance, free-space path loss, relative Tx--Rx angle, transmitter altitude, receiver altitude, and the empirical model prediction. The first five features describe the link geometric relationship and basic propagation conditions, while the empirical prediction serves as a physics-prior input. The classic close-in free-space reference distance (CI) path loss model is used as the empirical model \cite{7999294}:

    {\begin{multline}
    	\label{CI}
    	\mathrm{PL}^\mathrm{CI}(f_c, d_\mathrm{3D}) = \mathrm{FSPL}(f_c, d_0) + 10n\log_{10}\left(\frac{d_\mathrm{3D}}{d_0}\right),
    \end{multline}}

    \noindent where $n$ is the path loss exponent (PLE), $d_{\mathrm{3D}}$ is the Euclidean distance between Tx and Rx, and $\mathrm{FSPL}(f_c, d_0)$ is the free-space path loss at frequency $f_c$ and reference distance $d_0$, which is set to 1 m in this paper. By fitting the CI model on the training dataset, the PLE is obtained as $n=3.394$. The system branch employs a multilayer perceptron (MLP) to produce a 256-dimensional system feature vector. The image feature and the system feature are then organized as a joint token sequence. This dual-branch design enables the model to exploit both environmental image features and system parameters for path loss prediction.
    
	A feature fusion module based on multi-head self-attention (MHSA) is further introduced. Given the joint token sequence from the dual-branch feature extraction module, MHSA learns adaptive dependencies between the image features and the system features. The attended representations are then aggregated by an MLP to produce a 64-dimensional fused feature for subsequent prediction. Compared with direct concatenation, this design strengthens collaborative modeling between image features and system features.
	
	For the environmental compensation prediction module, we design an environmental compensation strategy to achieve deeper integration between the empirical model and the neural network. Conventional model-assisted approaches typically learn a compensation term on top of a fixed empirical model, such as the CI model or the 3GPP TR 38.901 large scale path loss model. In this setting, the empirical model imposes a single global distance trend, and the learned compensation mainly performs correction around that trend. In complex suburban environments, this limited form of correction may not provide sufficient flexibility, and the utilization of environmental information can remain limited. Motivated by the idea of piecewise modeling in statistical path loss analysis, where different propagation conditions are often characterized by different model parameters, we introduce an environment-adaptive adjustment of the CI model trend. In the CI model, the path loss exponent determines the distance variation trend under a given environment. Therefore, our network predicts a link-specific effective PLE from the fused features, and simultaneously predicts an additive compensation value. The final path loss prediction is obtained by combining the CI prediction with the learned compensation:
	
	\begin{equation}
		\hat{\mathrm{PL}}(f_c,d_{\mathrm{3D}})
		=\mathrm{FSPL}(f_c,d_0)
		+10\hat{n}_{\mathrm{eff}}\log_{10}\left(\frac{d_{\mathrm{3D}}}{d_0}\right)
		+\Delta\hat{\mathrm{PL}},
		\label{eq:pl_final}
	\end{equation}
	 
	 \noindent where $\hat{n}_{\mathrm{eff}}$ is the predicted link-specific effective PLE and $\Delta \hat{\mathrm{PL}}$ is the predicted compensation term. Different from the conventional global PLE in the CI model, $\hat{n}_{\mathrm{eff}}$ should be interpreted as an effective link-level parameter. Notably, $\hat{n}_{\mathrm{eff}}$ and $\Delta \hat{\mathrm{PL}}$ are not directly supervised during training. The loss function is computed only between the predicted path loss $\hat{\mathrm{PL}}$ and the ground-truth measurement, so the effective PLE is implicitly learned through end-to-end optimization. To ensure physical plausibility, $\hat{n}_{\mathrm{eff}}$ is produced by an MLP head and constrained to be non-negative using a ReLU activation.

		\begin{table}[!htbp]
		\centering
		\caption{Key Configurations of the Proposed Model Architecture}
		\renewcommand{\arraystretch}{1.5}
		
		\begin{tabular}{|c|c|c|}
			\hline
			Module & Component & Configuration \\
			\hline
			
			\multirow{2}{*}{\shortstack{Dual-branch\\feature extraction}}
			& System parameter MLP & $6 \rightarrow 128 \rightarrow 256$ \\
			\cline{2-3}
			& Image feature encoder & ResNet 50 $\rightarrow 256$ \\
			\hline
			
			\multirow{2}{*}{Feature fusion}
			& Fusion MHSA & Heads = 8 \\
			\cline{2-3}
			& Fusion MLP & $256 \rightarrow 128 \rightarrow 64$ \\
			\hline
			
			\multirow{2}{*}{Prediction head}
			& Effective PLE MLP & $64 \rightarrow 32 \rightarrow 1$ \\
			\cline{2-3}
			& Compensation MLP & $64 \rightarrow 32 \rightarrow 1$ \\
			\hline
			
		\end{tabular}
		\label{model_param}
	\end{table}
	
	In model implementation and training, path loss prediction is formulated as a regression task, and the root mean square error (RMSE) is adopted as the training loss and the primary evaluation metric. The system-parameter inputs are normalized using Min-Max normalization, while both satellite images and elevation maps are standardized using Z-score normalization. The key model settings are summarized in Table \ref{model_param}.  Adam is used as the optimizer with a learning rate of $1\times10^{-4}$, a batch size of 64, and 300 training epochs.

	\section{Performance Evaluation}
	
	To verify the effectiveness of different environmental image features for suburban path loss prediction, we first conduct comparative experiments under three environmental image formats, and evaluate two feature fusion schemes for combining system features and image features, including direct concatenation and the fusion module. To examine the impact of model capacity on performance, ResNet 34 and ResNet 50 are adopted as the image feature extraction backbones. It should be noted that this stage uses an end-to-end regression setting only, where the path loss value is directly predicted after feature extraction and fusion, without introducing any model-assisted component. This setting enables a fair assessment of how different environmental representations affect prediction performance and helps determine a suitable network size. 
	
	\begin{figure}[!b]
		\centering
		\includegraphics[width=3in]{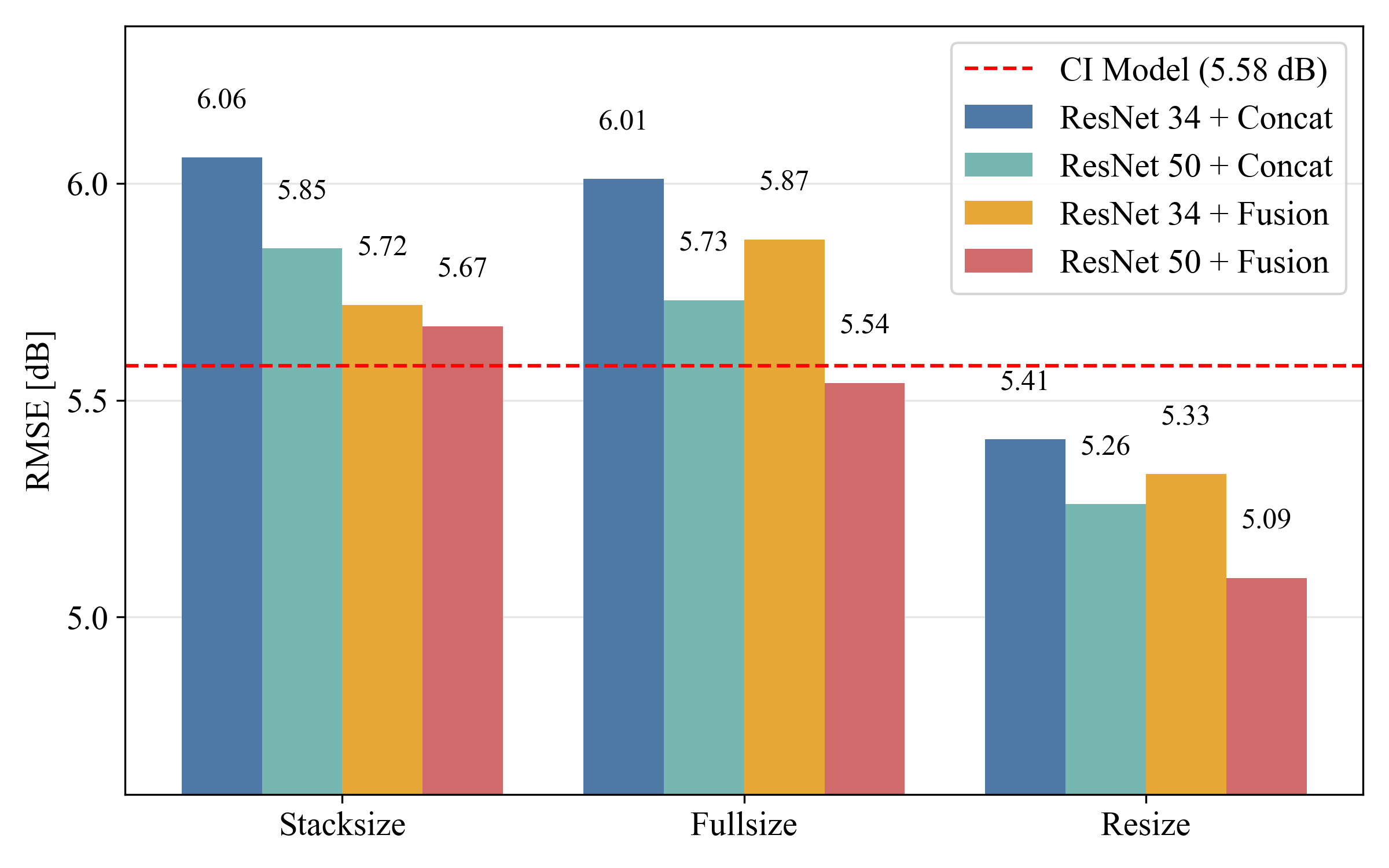}
		\caption{Test RMSE comparison under different environmental image formats and baseline model configurations.}
		\label{baseline_bar}
	\end{figure}
	
	As shown in Fig. \ref{baseline_bar}, the RMSE on the Resize dataset is overall lower than that on the Stacksize and Fullsize datasets. In the considered suburban scenario, broader environmental context may be more informative than link-centered geometry. Under the same setting, ResNet 50 generally performs better than ResNet 34. Further comparison between the two fusion schemes shows that, relative to direct concatenation, the proposed fusion module reduces the RMSE in most settings, suggesting that dynamic feature interaction helps strengthen the collaborative modeling between multimodal image features and system features. Overall, the configuration using the Resize representation, ResNet 50, and the fusion module achieves the best baseline performance, with an RMSE of 5.09 dB.

	\begin{table}[!t]
		\centering
		\caption{Test Set Performance and Ablation Comparison}
		\renewcommand{\arraystretch}{1.4}
		\setlength{\tabcolsep}{5pt}
		\begin{tabular}{l c c c}
			\toprule[1pt]
			Model & RMSE (dB) & MAPE (\%) & PCC \\
			\midrule
			Baseline model & 5.09 & 3.35 & 0.89 \\
			Distance-enhanced baseline & 4.98 & 3.17 & 0.90 \\
			CI model & 5.58 & 3.63 & 0.87 \\
			Model-assisted model & 4.77 & 3.04 & 0.91 \\
			Effective-PLE-only model & 4.88 & 3.31 & 0.91 \\
			Proposed model & 4.04 & 2.57 & 0.93 \\
			\bottomrule[1pt]
		\end{tabular}
		\label{tab:overall_perf}
	\end{table}

	Based on the first-stage comparison, the combination of ResNet 50 and the fusion module is used as the baseline model. Several comparison models are then evaluated, as summarized in Table \ref{tab:overall_perf}. The model-assisted model represents the conventional strategy that predicts only an additive compensation term on top of the CI model. The distance-enhanced baseline uses the same multimodal network as the baseline model, but concatenates the Tx--Rx distance with the fused representation before the prediction head. The effective-PLE-only model predicts the link-specific effective PLE while removing the additive compensation term.

	The proposed method achieves the best performance on all three metrics. Compared with the baseline model, the RMSE decreases from 5.09 dB to 4.04 dB, the mean absolute percentage error (MAPE) decreases from 3.35 percent to 2.57 percent, and the Pearson correlation coefficient (PCC) increases from 0.89 to 0.93. 	The distance-enhanced baseline achieves a slight improvement over the baseline model, which indicates that the performance gain cannot be attributed merely to stronger distance-related feature representation.  The proposed model outperforms both ablation models, suggesting that predicting the effective PLE and the additive compensation term provides a more effective model-assisted prediction strategy.
	
	\begin{figure}[!t]
		\centering
		\includegraphics[width=3in]{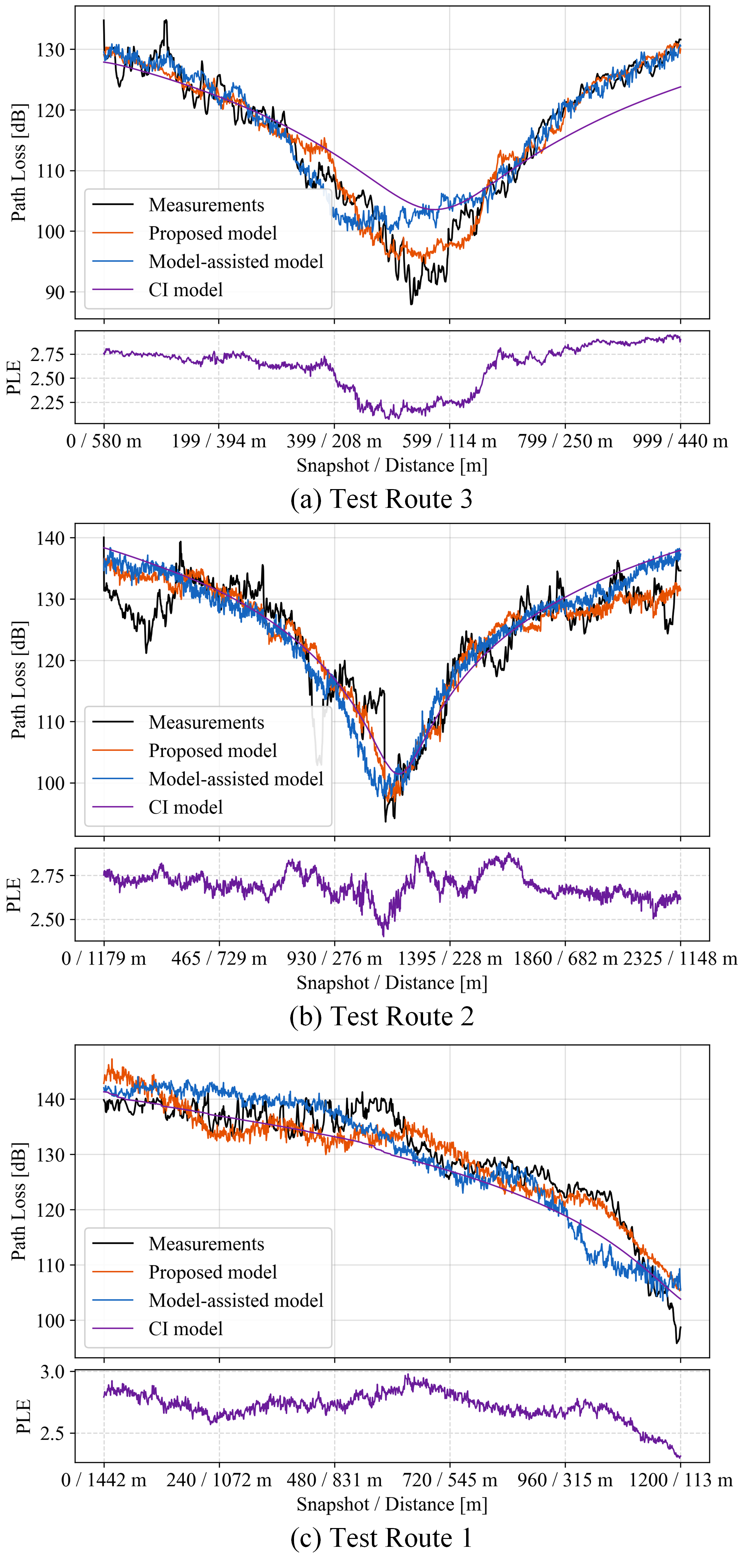}
		\caption{Path loss predictions and link-specific effective PLE trends on the three test routes in Fig. \ref{overview}. The upper and lower subplots show the path loss comparison and predicted effective PLE, respectively.}
		\label{2222}
	\end{figure}
	
	Fig. \ref{2222} further shows the predicted path loss curves on the three test routes labeled in Fig. \ref{overview}, which exhibit different path loss variation patterns in the suburban measurement area. The proposed method shows closer agreement with the measurements in both overall trend and local variations, which confirms the effectiveness of the proposed model-assisted prediction strategy in complex suburban scenarios. In addition, Fig. \ref{2222} presents the predicted effective PLE along the test routes. Although the effective PLE is not explicitly supervised during training, its values remain within a reasonable range and vary along the routes, indicating that the model can adaptively adjust the link-specific effective PLE according to the environmental and system features.

	\section{Conclusion}
	
	This paper investigates suburban path loss prediction using measurement data from Pingtan, China. Comparisons among three environmental image organizations show that the Resize representation is more effective in the considered suburban scenario. Based on this observation, we propose a model-assisted environmental compensation method that fuses environmental image with system features and jointly predicts a link-specific effective path loss exponent and a compensation term to adapt the empirical prior. Results show that the proposed method outperforms the empirical baseline,  conventional model-assisted method, and selected baseline model in both error and correlation metrics, achieving an RMSE of 4.04 dB, a MAPE of 2.57 percent, and a PCC of 0.93.

	\bibliographystyle{IEEEtran}
	
    \makeatletter
    \let\origbibitem\bibitem
    \renewcommand{\bibitem}[1]{%
      \normalcolor
      \origbibitem{#1}}
    \makeatother
	\bibliography{ref}
	
\end{document}